\documentclass[aps,prb,twocolumn,superscriptaddress, show pacs]{revtex4-2}
\usepackage{bm, amsmath}
\usepackage{graphicx}
\usepackage{color}
\usepackage{epstopdf}
\usepackage{graphicx}
\usepackage{dcolumn}
\usepackage{bm}
\usepackage{subfigure}
\usepackage[rawfloats=true]{floatrow} 
\begin{document}

\title{Communicating skyrmions as the main mechanism underlying skyrmionium (meta)stability in quasi-two-dimensional chiral magnets}

\author{Kaito Nakamura}
\affiliation{Department of Chemistry, Faculty of Science, Hiroshima University Kagamiyama, Higashi Hiroshima, Hiroshima 739-8526, Japan}

\author{Andrey O. Leonov}
\thanks{Corresponding author: leonov@hiroshima-u.ac.jp}
\affiliation{Department of Chemistry, Faculty of Science, Hiroshima University Kagamiyama, Higashi Hiroshima, Hiroshima 739-8526, Japan}
\affiliation{International Institute for Sustainability with Knotted Chiral Meta Matter, Kagamiyama, Higashi Hiroshima, Hiroshima 739-8526, Japan}

\date{\today}

\begin{abstract}
{We re-examine the internal structure of skyrmioniums stabilized in quasi-two-dimensional chiral magnets with easy-axis uniaxial anisotropy. Skyrmioniums are particle-like states of two nested skyrmions with opposite polarities contributing to zero topological charge. The physical principles of 
skyrmionium stability are drawn from both the analytical analysis with a trial function 
and from numerical simulations within the framework of micromagnetism.  We deduce that the radii of the internal skyrmion with the positive polarity and the ring-shaped external skyrmion with the negative polarity are mutually dependent, which constitutes the paradigm of communicating skyrmions. For large central skyrmions, the skyrmionium transforms into a narrow circular domain wall, whereas for small internal radii, the ring expands, which occurs at the verge of collapsing into an ordinary isolated skyrmion. We show that skyrmioniums may form lattices of two varieties depending on the polarity of the internal skyrmion. At the phase diagram (magnetic field)-(uniaxial anisotropy), both skyrmionium lattices share the same area with one-dimensional spiral states  and remain metastable solutions for the whole range of control parameters. By expanding at the critical line, skyrmionium lattices do not release isolated skyrmioniums. Isolated skyrmioniums of just one type exist apart from the corresponding lattice in a narrow field region restricted by the critical line of expansion from below and by the line of collapse above. 
}
\end{abstract}

         
\maketitle

\section{Introduction}

Chiral magnetic skyrmions \cite{JETP89,NT} are topological solitons surrounded by the homogeneously magnetized  states and stabilized by Dzyaloshinskii-Moriya interaction (DMI) \cite{Moriya,Dz58}. 
Their characteristic length scale \cite{Bogdanov94,Bogdanov99,Le} results from the competition between exchange interaction and DMI and ranges from few atomic spacings up to microns \cite{Wiesendanger2016}.
Skyrmions were first experimentally observed in bulk cubic helimagnets (MnSi \cite{Muehlbauer09} and FeGe \cite{FeGe}, and the Mott insulator, Cu${_2}$OSeO${_3}$ \cite{Seki2012}) where they represent three-dimensional (3D) tubes along the field direction \cite{Damien,Birch:2020}. 
Subsequently, 3D isolated skyrmions (IS) \cite{twists1,twists2} have been microscopically studied in thin layers of cubic helimagnets (Fe,Co)Si \cite{yuFeCoSi} and FeGe \cite{yuFeGe} where they gain stability in a broad range of temperatures and magnetic fields \cite{twists}.

Essentially two-dimensional (2D) skyrmions are stabilized, e.g., in bulk polar magnets with C$_{nv}$ symmetry, such as  GaV$_4$S$_8$ and GaV$_4$Se$_8$ \cite{Bordacs17,Fujima17}, in which skyrmions propagate into the third direction without modifying their 2D pattern. 
In these N\'eel skyrmions, the magnetization rotates along the radius-vector from the skyrmion center to the outskirt. 
On the other hand, thin film multilayer structures open up prospects of manipulating skyrmions on a 2D arena. 
The inversion symmetry breaking and the induced DMI originate from the interfaces between a heavy metal layer and the skyrmion-hosting magnetic layer as occurs, e.g., in  PdFe/Ir (111) bilayers \cite{Romming2013}. Such systems are extremely versatile as for the choice of the magnetic, non-magnetic and capping layers in addition to the possibility to be stacked.

Nowadays, skyrmions attract  enormous interest due to the perspectives of their applications in information storage and processing devices \cite{Sampaio13,Tomasello14,Shigenaga}.
Indeed, skyrmions are topologically stable \cite{Cortes-Ortuno}, they have the nanometer size \cite{Wiesendanger2016} and can be manipulated by electric currents \cite{Schulz,Jonientz} of small densities.  In particular, in the skyrmion racetrack \cite{Wang16,Fert2013}, information flow is encoded in isolated skyrmions driven along  a narrow strip.
At the same time, there is an obvious obstacle towards the practical use of skyrmion-based devices -- the skyrmion Hall effect (SHE), which leads to the  curved trajectory of moving skyrmions \cite{Toscano,Gobel}.
The main strategy to overcome this obstacle is to consider skyrmion-based solitons with zero topological charge, which would be able to cancel the Magnus force. Among such skyrmion varieties are  antiferromagnetic skyrmions \cite{AFM}, states of coupled merons with the opposite topological charges \cite{Mukai} and/or target-skyrmions \cite{Leonov14}.

Originally, 2D target-skyrmions were introduced in Ref. \cite{Bogdanov99} under the name $k\pi$-skyrmions. They consist of  a central skyrmion with either polarity and a number of concentric helicoidal undulations: the magnetization rotates by an angle $k\pi$ between the center and the surrounding ferromagnetic state (with $k$ integer $>0$).    The topological charge alternates between 1 or 0 depending whether $k$ is odd or even. 
Skyrmionium ($2\pi$-skyrmion) represents a second member of the $k\pi$-skyrmion family \cite{Komineas},  which has a topological charge $Q = 0$ and is thus bound to avert SHE. Among other advantages of skyrmioniums over ordinary $\pi$-skyrmions is higher mobility \cite{Kolesnikov18}, which makes them a good alternative for spintronic devices \cite{Wang20}. So far, the crucial question of skyrmionium stability was addressed by computing the energy barrier with respect to the ordinary skyrmion through the geodesic nudged elastic band method \cite{Hagemeister}. Thermal annihilation of skyrmioniums and their transformation into a skyrmion was studied in Ref. \cite{Jiang24} by analytical and numerical methods of micromagnetism.

High-symmetry nanostructured objects (like magnetic nanowires \cite{Higgins}, nanodisks \cite{Butenko}, or nanorings \cite{Ponsudana21}) are commonly used systems to host target-skyrmions since they provide  the stabilization effect of surfaces and supplement the target-skyrmions with additional negative energy from the edge states, which may even favor them over other solitons \cite{Leonov14}. 
Recently, target-skyrmions were generated by weakly coupling 30nm thin Permalloy (Ni$_{80}$Fe$_{20}$) disks with a 1 $\mu$m diameter to asymmetric (Ir 1 nm/Co 1.5 nm/Pt 1 nm) $\times$ 7 multilayers that exhibit Dzyaloshinskii-Moriya interaction \cite{Kent}. Off-axis electron holography was used  to record images of target-skyrmions in a 160-nm-diameter nanodisk of the chiral magnet FeGe \cite{Zheng}.
Skyrmioniums have also been experimentally spotted in thin-film geometries lacking the stabilization support from the side boundaries. 
The observation of skyrmioniums in thin ferromagnetic films coupled to a magnetic topological insulator was reported in Ref. \cite{Zhang}. Skyrmioniums have also been investigated in a frustrated Kagome magnet Fe$_3$Sn$_2$ \cite{Yang23} and in flakes of the van der Waals magnet Fe$_{3-x}$GeTe$_2$ \cite{Pwoalla23}.

In the present paper, to address the problem of skyrmionium stability with the simultaneous effect of an applied magnetic field and an easy-axis anisotropy, 
we re-examine the internal structure of skyrmioniums by using the linear Ansatz and numerically rigorous solutions. 
We show that skyrmionium represents a pair of communicating skyrmions: whereas the central skyrmion aspires to adapt the magnetization rotation based on the same energy arguments as for an ordinary skyrmion, the ring-shaped external skyrmion contracts in an attempt to reduce its own surface area. The ratio of radii of communicating skyrmions is balanced to reach a local energy minimum: for large radii of the central skyrmion, the skyrmionium transforms into a circular domain wall with the comparable radius of the surrounding ring; for small radii of the central skyrmion, the skyrmionium increases the radius of the external skyrmion but still inevitably transforms into a skyrmion as was discussed in Refs. \cite{Hagemeister,Jiang24}.

We also address the field- and anisotropy-driven transformations of skyrmionium lattices (SkmL) of two varieties depending on the polarity of the central skyrmion. 
At the phase diagram of states in coordinates (magnetic field)-(uniaxial anisotropy), the skyrmionium lattices were found to occupy the same region as spiral states but with higher energy what makes them only metastable solutions. 
On the contrary to skyrmion lattices, no isolated skyrmioniums were released during the transition of skyrmionium lattices into the homogeneous state. Isolated skyrmioniums exist as a separate branch of solutions within the narrow area on the phase diagram restricted by the lines of their collapse or expansion.

\begin{figure*}
\includegraphics[width=1.99\columnwidth]{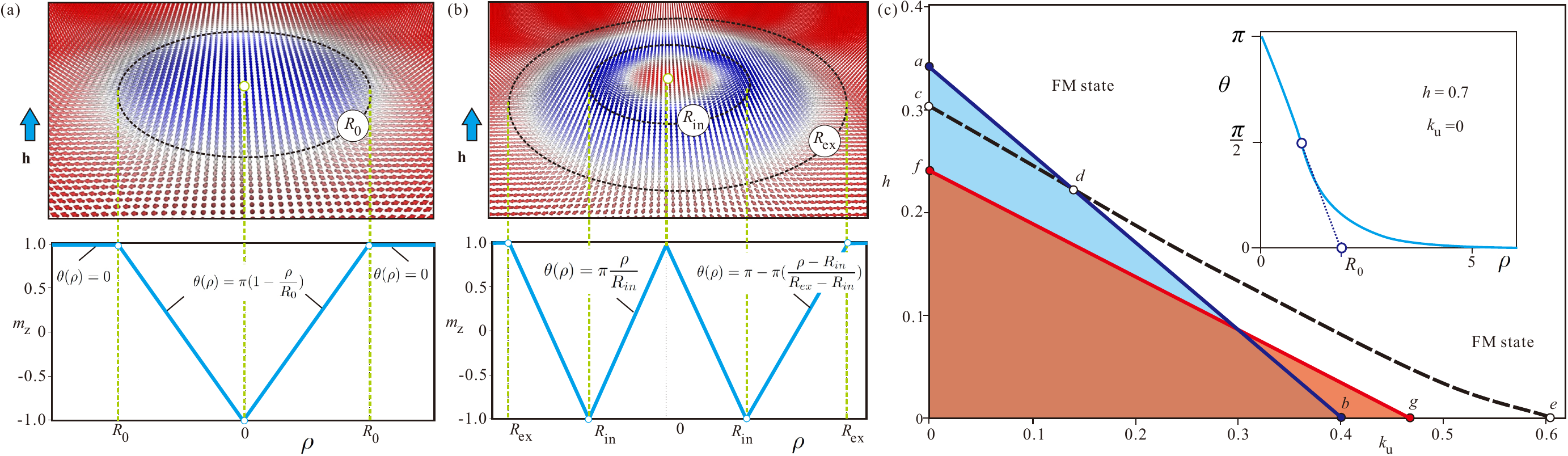}
\caption{(color online) (a), (b) Schematics of isolated N\'eel skyrmions and skyrmioniums in polar magnets with C$_{nv}$ symmetry (or in multilayers with the induced DMI). As trial functions for angular skyrmion/skyrmionium profiles, linear Ansatzes are used (blue lines in the lower panels). 
(c) The simplified diagram on the plane $(k_u,h)$ constructed for the trial functions. In the red- and blue-shaded regions, skyrmioniums and skyrmions acquire the negative eigen-energy with respect to the homogeneous state, which signifies their condensation into a lattice. The inset shows an example of an arrow-like angular profile for a skyrmion with $h=0.7, k_u=0$, which justifies the used linear Ansatzes.  
\label{fig:Ansatz}}
\end{figure*}

\section{Phenomenological theory of skyrmioniums in two-dimensional helimagnets \label{section:phenomenology}}

\subsection{Micromagnetic energy functional}

The magnetic energy density of a chiral magnet with C$_{nv}$ symmetry can be written as the sum of the exchange, the DMI, Zeeman, and the anisotropy
energy densities, correspondingly:
\begin{equation}
w(\mathbf{m})=\sum_{i,j}(\partial_i m_j)^2+w_{DMI}-\mathbf{m}\cdot\mathbf{h} - k_u m_z^2.
\label{functional}
\end{equation}
Spatial coordinates $\mathbf{x}$ are measured in units of the characteristic length of modulated states $L_D=A_{ex}/D_{DMI}$. $A_{ex}>0$ is the exchange stiffness, $D_{DMI}$ is the Dzyaloshinskii constant. $k_u=K_uM^2A_{ex}/D_{DMI}^2$ is the non-dimensional anisotropy constant. We restrict ourselves by the easy-axis case, i.e., $k_u>0$.

In the following, we consider a thin film of a ferromagnetic material on the $xy$-plane using periodic boundary conditions. $\mathbf{h} = \mathbf{H}/H_0$ is the  magnetic field applied along $z$ axis, $H_0 = D_{DMI}^2/A_{ex}|\mathbf{M}|$.
The magnetization vector $\mathbf{m}(x,y) = \mathbf{M}/|\mathbf{M}|$ has a fixed length normalized to unity.
The DMI energy density  has the following form specific for magnets with the C$_{nv}$ symmetry:
\begin{equation}
w_{DMI} = m_x\partial_x m_z-m_z\partial_x m_x+m_y\partial_y m_z-m_z\partial_y m_y, 
\label{DMI}
\end{equation}
where $\partial_x=\partial/\partial x,\, \partial_y=\partial/\partial y$.

The phase diagram of states for model (\ref{functional}) on the plane of control parameters $h$ and $k_u$ was plotted, e.g., in Ref. \cite{Mukai} and features only modulated one-dimensional 1D (spirals) and two-dimensional 2D (skyrmions) phases with the propagation directions perpendicular to the polar axis. A vast area of the phase diagram is occupied by the field-polarized homogeneous state, which may host isolated solitons such as kinks, skyrmions and/or skyrmioniums. 1D kinks and 2D skyrmions in this area can be an outcome of a gradual expansion of the spiral state and/or the skyrmion lattice. And vice versa, isolated entities may condense into extended modulated states 
when the eigen-energy of an isolated soliton becomes negative with respect to the surrounding homogeneous state and the solitons tend to fill the whole space with some equilibrium intersoliton distance. The mechanism of modulated phase formation through nucleation and condensation of isolated solitons follows a classification introduced by de Gennes \cite{deGennes} for (continuous) transitions into incommensurate modulated phases. The transformation of skyrmions during the formation of the lattice was first investigated in Ref. \cite{Bogdanov94}. In the present paper, however, it will be shown that this scenario is not the case for skyrmioniums.

As a primary numerical tool to minimize the functional (\ref{functional}), we use MuMax3 software package (version 3.10) which calculates magnetization dynamics by solving the Landau-Lifshitz equation using finite difference discretization technique \cite{mumax3}.
To double-check the validity of obtained solutions, we also use our own numerical routines, which are explicitly described in, e.g., Ref. \onlinecite{metamorphoses} and hence will be omitted here. 

Since isolated skyrmions and skyrmioniums are axi-symmetric particle-like states, we will use the spherical coordinates for the magnetization: 
\begin{equation}
\mathbf{m}= (\sin \theta(\rho) \cos \psi(\varphi),\sin \theta(\rho) \sin \psi(\varphi), \cos \theta(\rho)),
\label{SphericalM}
\end{equation}
and cylindrical coordinates for the spatial variables \cite{Bogdanov94,Bogdanov99}
\begin{equation}
\mathbf{r}=(\rho \cos \varphi, \rho \sin \varphi,z)
\label{Cylindrical}
\end{equation}
where $\psi(\varphi)=\varphi$ is adapted for the N\'eel-like magnetization rotation.

The total energy of an isolated skyrmionium with respect to the homogeneous state can be written as
\begin{align}
E = &\int\limits_0^{\infty} \varepsilon (\theta, \rho) 2\pi\rho d \rho,\, \nonumber\\
\varepsilon (\theta, \rho) =  &[\left(\frac{d \theta}{d \rho } \right)^2 +\frac{\sin^2 \theta}{\rho^2} 
+ h\,(1- \cos \theta) + \frac{d \theta}{d \rho } + \nonumber\\
&+ \frac{\sin 2 \theta}{2\rho}+k_u(1-\cos^2 \theta)]
\label{energy1}
\end{align}
where $\varepsilon (\theta, \rho)$ is an  energy density.

\begin{figure*}
\includegraphics[width=1.99\columnwidth]{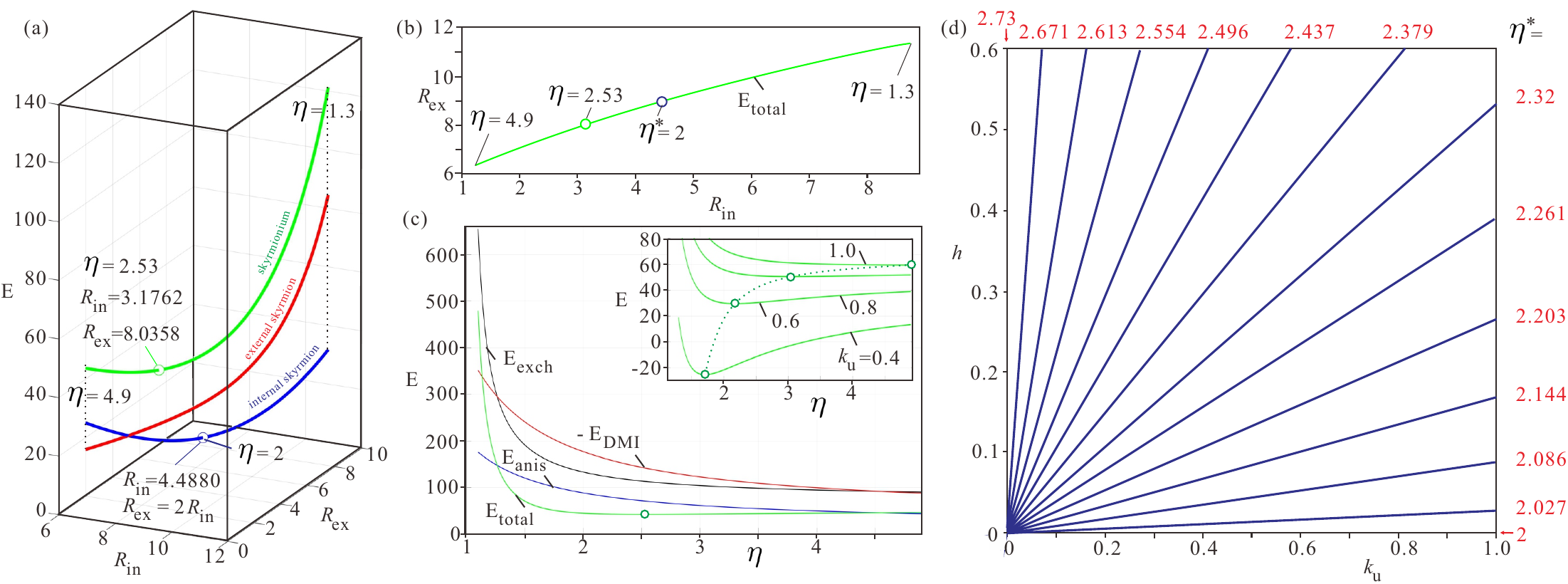}
\caption{(color online) (a) The total energy of an isolated skyrmionium plotted in dependence on the radii of the internal $R_{in}$ and the external $R_{ex}$ skyrmions (green curve). For the fixed radius $R_{in}$, the radius $R_{ex}$ is defined by the ratio $\eta$, which changes along the energy curve from the high value at the right to unity at the left. The total energies of the internal and the external skyrmions are shown by blue and red curves, correspondingly. The energy minimum is reached at $R_{in}=3.17$ and $\eta=2.53$ for $h=0, k_u=0.7$. (b) The corresponding energy curve viewed from above. (c) The separate energy contributions depending on the ratio $\eta$: black curve -- the exchange energy; red curve -- the DMI energy; blue curve -- the anisotropy energy. The energy minimum at the green curve is highlighted by the circle.   The inset shows the total energy of a skyrmionium for different values of the uniaxial anisotropy, which may become negative or may loose its minimum. (d) The lines on the plane $(k_u,h)$, along which the energy minimum is reached for an internal skyrmion only.
\label{fig:02}}
\end{figure*}

\section{Analytical results for the linear Ansatz}

\subsection{Isolated skyrmions}

In a wide range of control parameters $k_u$ and $h$, skyrmion profiles $\theta(\rho)$ are known to bear strongly localized character (see, e.g., an inset in Fig. \ref{fig:Ansatz} (c) plotted for $h=0.7, k_u=0$). 
According to the conventions of Refs. \cite{Bogdanov94,Bogdanov99,pss94} such  arrow-like solutions  can be decomposed into skyrmionic cores with linear dependence 
\begin{equation}
\theta(\rho)=\pi(1-\frac{\rho}{R_0}),\, \rho \leq L_D 
\label{linearIS}
\end{equation}
and exponential "tails" with 
\begin{equation}
\theta\propto \exp{[-\alpha\rho]},\,\rho \gg L_D .
\label{exponenta}
\end{equation}

Therefore, a "nucleus" with a diameter $2R_0$ can be considered as a two-dimensional particle-like state as it accumulates almost all energy of the isolated skyrmion.  At the same time the asymptotic exponential tails are viewed as the "field" generated by the particle \cite{JETP95}. 

In the following, we focus on the physical principles drawn from the linear Ansatz (\ref{linearIS}) with $\rho\leq R_0$ and the condition $\theta(\rho)=0, \rho>R_0$ (Fig. \ref{fig:Ansatz} (a)).
Equilibrium radius $R_0$ of the skyrmion core can be found from  substituting the linear Ansatz into (\ref{energy1}), integrating, and then minimizing with respect to $R_0$.
Thus, the skyrmion energy (\ref{energy1}) is reduced to a quadratic potential
\begin{equation}
E (R_0)  = D_0 -2 A_0 R_0  + (B_0 h+C_0k_u) R_0^2,
\label{lineEnergy}
\end{equation}
with parameters 
\begin{align}
&A_0=\int_0^{R_0} [\frac{d \theta}{d \rho } + \frac{\sin 2 \theta}{2\rho}]2\pi\rho d\rho=4.9348,\nonumber\\
&B_0=\int_0^{R_0} [(1- \cos \theta)]2\pi\rho d\rho=1.86835,\nonumber\\
&C_0=\int_0^{R_0} [(1-\cos^2 \theta)]2\pi\rho d\rho=\pi/2,\nonumber\\
&D_0=\int_0^{R_0} [\left(\frac{d \theta}{d \rho } \right)^2 +\frac{\sin^2 \theta}{\rho^2} ]2\pi\rho d\rho=38.6644.
\label{constants0}
\end{align}

The equilibrium value of the skyrmion radius 
\begin{equation}
R_0^{min} = \frac{A_0}{(B_0 h+C_0k_u)}.
\label{lineRIS}
\end{equation}
arises as a result of the competition between chiral and Zeeman/anisotropy energies.

The exchange energy $D_0$ does not depend on the skyrmion size and presents an amount of positive energy "trapped" within the skyrmion. Moreover, the size of IS diverges by approaching the critical point $(0,0)$ at the phase diagram.
In centrosymmetric systems without DMI, localized solutions are radially unstable and collapse spontaneously under the influence of applied magnetic field and anisotropy \cite{Bogdanov94}. Thus, already such a  simplified model offers an important insight into physical mechanisms underlying the formation of the chiral skyrmions.

At the phase diagram (Fig. \ref{fig:Ansatz} (c)), the energy of an isolated skyrmion (\ref{lineEnergy}) acquires the negative values with respect to the homogeneous state within the triangular region $a-b-0$. Remarkably, a small fraction of this region ($a-d-c$) oversteps the line $c-d-e$, at which the spiral state turns into a system of isolated kinks.  Based on this observation, in Ref. \cite{JETP89}, it was concluded that a skyrmion lattice would represent a thermodynamically stable state within some region of this phase diagram. 
Subsequent numerically rigorous calculations found a vast area of the SkL stability with the simultaneous effect of the applied magnetic field and the uniaxial anisotropy \cite{Bogdanov94,Mukai}.

\begin{figure*}
\includegraphics[width=1.5\columnwidth]{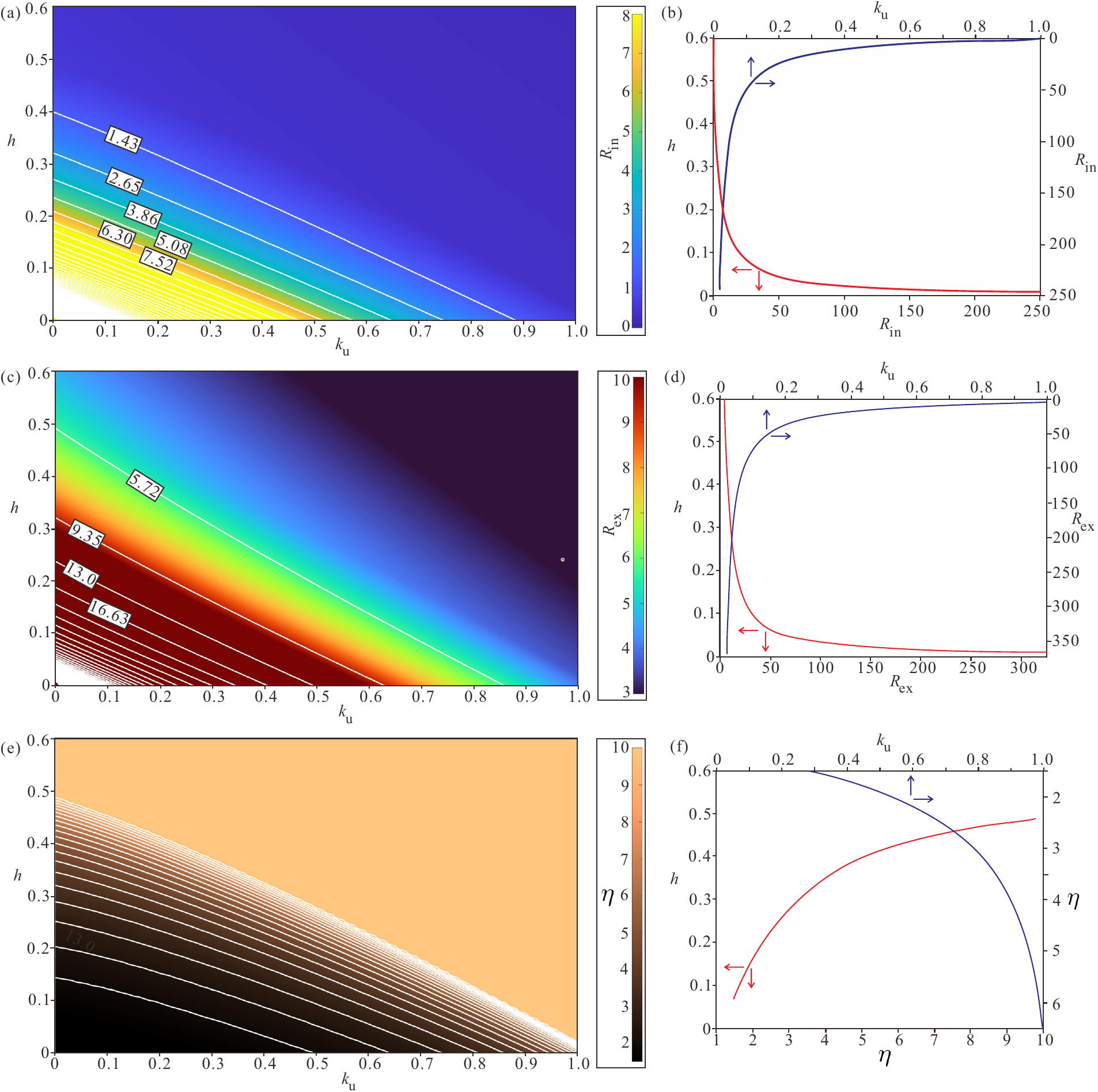}
\caption{(color online) Equilibrium parameters $R_{in}$ (a), $R_{ex}$ (c) and $\eta$ (e) corresponding to the energy minima on the plane $(k_u,h)$. White level lines and legends exhibit the exact values. 
The graphs reflect skyrmionium collapse at high magnetic fields and anisotropies, when the radii drastically decrease with the simultaneous growth of the ratio $\eta$. The right panels (b), (d), (f) show cross-cuts with zero anisotropy (blue curves) and/or zero magnetic field (red curves). 
\label{fig:03}}
\end{figure*}

\subsection{Isolated skyrmioniums \label{section:Skmlinear}}

In accordance with  the paradigm developed for ISs, we use the following trial function for skyrmioniums:
\begin{align}
&\theta(\rho)=\pi\frac{\rho}{R_{in}}, \rho\leq R_{in}, \nonumber\\
&\theta(\rho)=\pi-\pi(\frac{\rho-R_{in}}{R_{ex}-R_{in}}), R_{in}<\rho\leq R_{ex},\nonumber\\
&\theta=0, \rho>R_{ex}
\label{linearSkm}
\end{align}
where $R_{in}$ and $R_{ex}$ are the radii of the nested internal and the external ring-shaped skyrmions with the opposite polarities (Fig. \ref{fig:Ansatz} (b)).

By substituting this Ansatz into (\ref{energy1}), the equilibrium radii can be determined as:
\begin{equation}
R_{in} = \frac{A(\eta)}{B(\eta) h+C(\eta) k_u}, R_{ex}=\eta R_{in}. 
\label{lineRSkm}
\end{equation}
where the parameter $\eta$ denotes the ratio between two radii.
The total energy 
\begin{equation}
E(\eta)=D(\eta) -2 A(\eta) R_{in}(\eta)  + (B(\eta) h+C(\eta)k_u) R_{in}^2(\eta) \nonumber
\end{equation}
becomes an involved function of $\eta$. 
Each of the parameters $A(\eta), B(\eta), C(\eta), D(\eta)$ can be represented as a sum of the contributions from the internal ($\rho\leq R_{in}$) and the external ($R_{in}<\rho\leq R_{ex}$) skyrmions, e.g., $A(\eta)=A_{in}+A_{ex}(\eta)$:
\begin{align}
&A_{in}=A_0, A_{ex}(\eta)=A_0(1+\eta), \nonumber\\
&B_{in}=4.4148, B_{ex}(\eta)=(B_0\eta^2+2.547\eta-4.4148), \nonumber\\
&C_{in}=\pi/2, C_{ex}(\eta)=\pi\eta^2/2-\pi/2, \nonumber\\
&D_{in}=D_0, D_{ex}(\eta)=\frac{36.74\eta+21.28}{\eta-1}.
\label{constants}
\end{align}
The parameteres for the internal skyrmion are constants and equal the corresponding values for the isolated skyrmion (\ref{constants0}), except the Zeeman energy $B_{in}$. The integration occurs from $0$ to $R_{in}$ for the internal skyrmion, and from $R_{in}$ to $\eta R_{in}$ for the external one.

\begin{figure*}
\includegraphics[width=1.6\columnwidth]{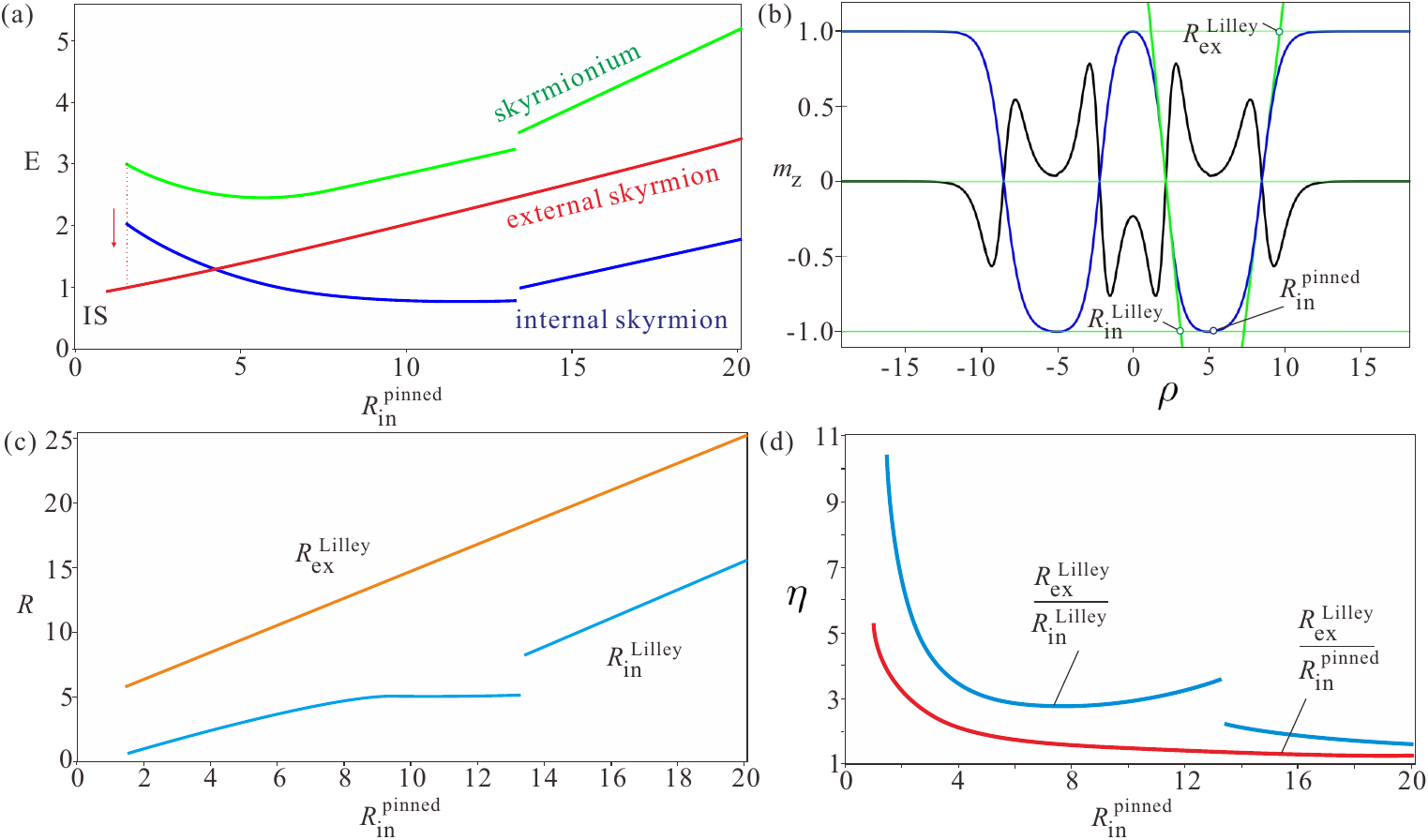}
\caption{(color online) (a) The total energy of an isolated skyrmionium (green curve) obtained by numerically minimizing the functional (\ref{functional}) with $h=0, k_u=0.7$. The total energies of the internal and the external skyrmions are shown by blue and red curves, correspondingly. During the numerical procedures, the magnetization vectors along the circle with the radius $R_{in}$ were pinned (see also text for details on the explanation of discontinuities at the energy curves). (b)  Definition of the radii $R_{in}$ and $R_{ex}$ according to the Lilley's convention. Blue profile represents $m_z$-component of the magnetization, black profile -- its second derivative. Green thick lines are tangents to the $m_z$ curve constructed at the points with zero second derivatives. Intersection of the tangent lines with the magnetization levels $\pm 1$ allows to introduce new characteristic radii $R_{in,ex}^{Lilley}$. (c) Evolution of the corresponding Lilley's radii for the skyrmionium-solutions obtained with MuMax3 for the pinned spins $m_z=-1$ around the circle with the radius $R_{in}^{pinned}$.  (d) The ratio of characteristic radii in dependence on $R_{in}^{pinned}$.
\label{fig:04}}
\end{figure*}

%

As an instructive example, we consider the skyrmionium solutions for fixed control parameters, $h=0, k_u=0.7$. 
Fig. \ref{fig:02} (a) shows the energies of the internal (blue line) and the external (red line) skyrmions as well as the total energy of a skyrmionium (green line)  in this case. 

First of all, we notice that all energies represent parametric curves of the ratio $\eta$, i.e., for a given radius $R_{in}$ of an internal skyrmion, the radius $R_{ex}$ of the external skyrmion is uniquely specified. Fig. \ref{fig:02} (b) shows the corresponding projection onto the plane $(R_{in},R_{ex})$ to additionally highlight the two-dimensional character of the energy dependence. 

In other words, both skyrmions "communicate" to reach the energy minimum for a fixed $R_{in}$: for small values of $R_{in}$ (close to the collapse of skyrmionium), the parameter $\eta$ acquires larger values (at the left side of the depicted energy curves in Fig. \ref{fig:02} (a), (b) $\eta=4.9$); 
for large values of $R_{in}$, the parameter $\eta$ tends to unity, which transforms a skyrmionium into a narrow circular domain wall (at the right side of the depicted energy curves $\eta=1.3$).

\begin{figure*}
\includegraphics[width=1.99\columnwidth]{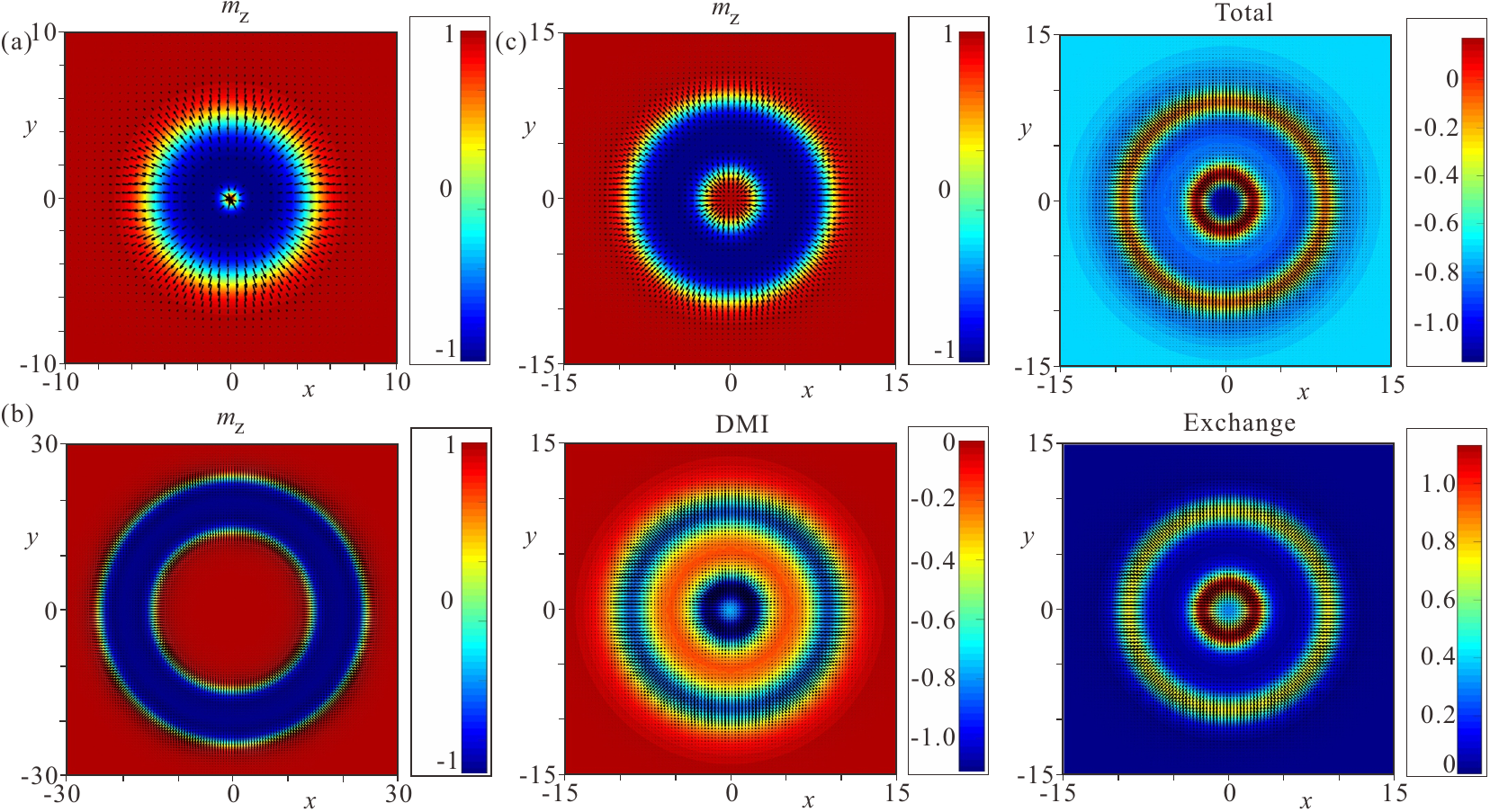}
\caption{(color online) Internal structure of skyrmioniums shown with the help of color plots of the $m_z$-component of the magnetization for limiting cases of large (a) and small (b) ratio $\eta$ as well as for the equilibrium solution corresponding to the energy minimum (c). $h=0, k_u=0.7$. Additional panels in (c) show color plots for different energy contributions.  
\label{fig:05}}
\end{figure*}

Second, only the energy of the internal skyrmion exhibits a minimum for some ratio  $\eta^*$, which represent straight lines on the plane $(k_u,h)$ (Fig. \ref{fig:02} (d)). We notice that  $\eta^*>2$, i.e., the internal skyrmion prefers "longer" rotation within the external skyrmion.   
For zero magnetic field, $\eta^*=2$, and the radius of the internal skyrmion equals the radius of the skyrmion, $R_{in}=R_0=4.488$, i.e., the internal skyrmion exhibits the magnetization  rotation according to the principles drawn in the previous subsection. For an applied magnetic field, however, the situation is different since the ordinary skyrmion has the negative polarity, whereas the internal skyrmion within the skyrmionium -- the positive one. 

The positive energy of the external ring-like skyrmion does not have any energy minimum (red curve in Fig. \ref{fig:Ansatz} (a)). However, it can be reduced by decreasing $R_{in}$ and by increasing $\eta$, i.e., the external skyrmion squeezes the internal one  in an attempt  to reduce its own surface area. 

The resulting skyrmionium is thus shaped as a compromise between these tendencies of communicating skyrmions:  the equilibrium internal radius in this case $R_{in}=3.1762$ is slightly smaller as compared with the skyrmion radius $R_0$; $\eta=2.53$. 

Third, the metastability of skyrmionium can also be considered from the energetic point of view. Fig. \ref{fig:02} (c) shows each energy contribution depending on the parameter $\eta$. 
The exchange energy 
diverges for $\eta\rightarrow 1$ and tends to 75.4 for $\eta\rightarrow\infty$. The exchange energy outweighs the DMI for small and large $\eta$. For moderate values of $\eta$, the DMI energy dominates and leads to some equilibrium characteristic size. Inset of Fig. \ref{fig:02} (c) shows that for small $k_u$ the energy of a skyrmionium may become negative, which facilitates condensation into a skyrmionium lattice. According to the phase diagram (Fig. \ref{fig:Ansatz} (c)), the hexagonal SkmL is formed 
below the line $f-g$, which is located deep inside the region of the spiral stability. One can provisionally  conclude that a skyrmionium lattice will remain a metastable state for the whole range of the control parameters.
For larger anisotropy values (inset of Fig. \ref{fig:02} (c)), the energy minimum shallows and with the condition $\eta\rightarrow\infty$ disappears, which is equivalent to the skyrmionium collapse. 

%

Fig. \ref{fig:03} shows the equilibrium parameters $R_{in}$ (a), $R_{ex}$ (c) and $\eta$ (e) on the plane $(k_u,h)$. These parameters are constant along almost straight lines.
Both the increasing magnetic field and the uniaxial anisotropy lead to the skyrmionium collapse, which is accompanied by the shrinking radii $R_{in}$ and $R_{ex}$, but the drastically increasing ratio $\eta$. 
%
%
Panels (b), (d), and (f) show two-dimensional cross-cuts either for zero anisotropy (blue curves) or zero magnetic field (red curves).
The linear Ansatz (\ref{linearSkm}) enables skyrmioniums with infinitely small internal radii (\ref{lineRSkm}), which precludes from plotting the exact line of skyrmionium collapse. Indeed, only for $\eta\rightarrow\infty$, $R_{in}\rightarrow 0$.

\begin{figure*}
\includegraphics[width=1.9\columnwidth]{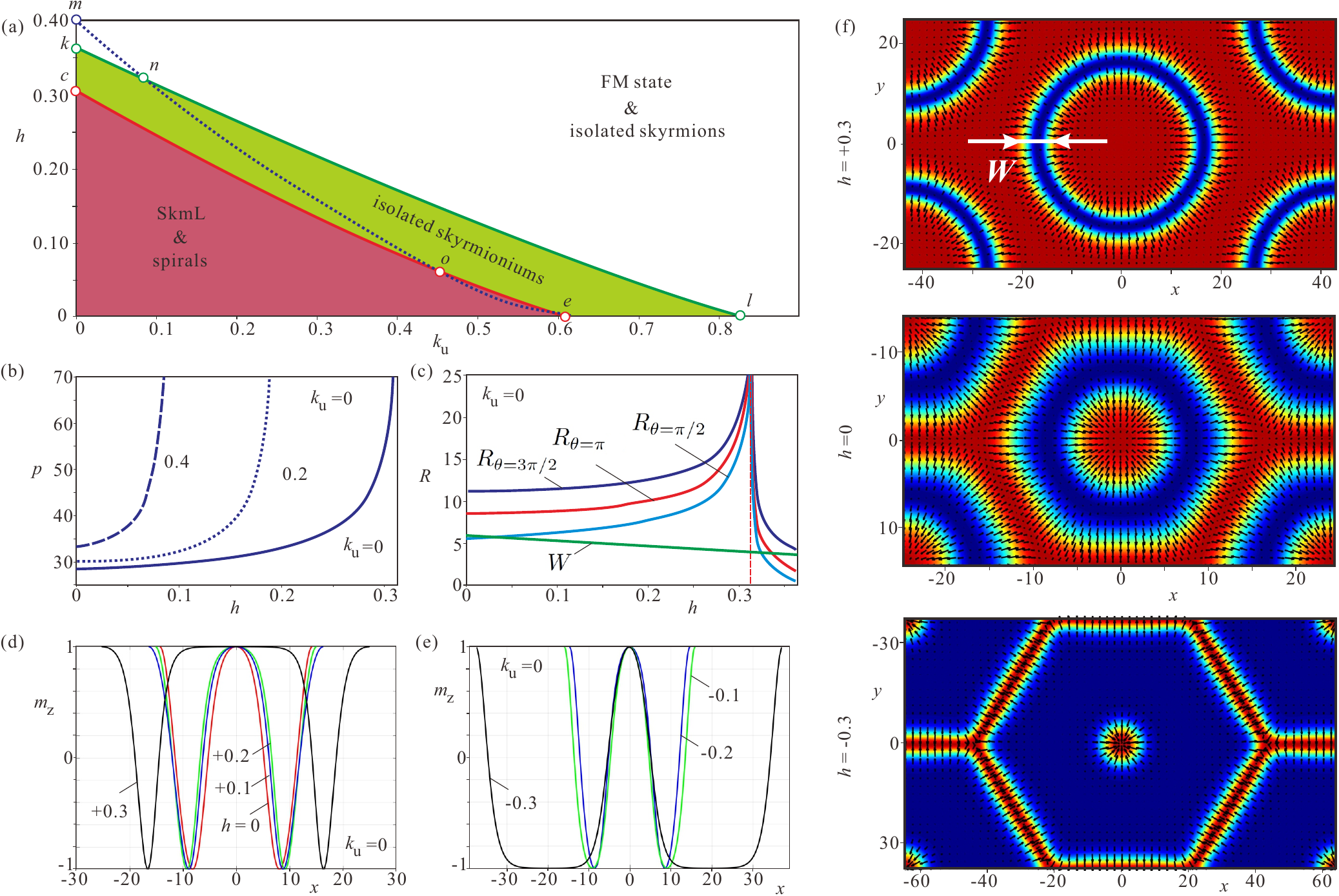}
\caption{(color online) (a) The abridged phase diagram for model (\ref{functional}) on the plane of control parameters $(k_u,h)$. Red-shaded area is occupied by the spiral states and the metastable SkmLs. At the critical line $c-d$, both types of SkmLs expand. The diverging period of a SkmL for different values of the uniaxial anisotorpy is plotted in (b). The green-shaded area accommodates isolated skyrmions, which also inflate at the bottom line $c-e$ and collapse at the line $k-l$. The dotted line $m-n-o-e$ shows the corresponding characteristic line for SkL, at which the SkL expands and releases isolated skyrmions. (c) Field-driven evolution of other characteristic radii at the levels of the magnetization with $\theta=\pi/2, \pi, 3\pi/2$ for both SkmL and isolated skyrmioniums. All radii diverge at the critical line $c - e$. The width of the circular domain wall, however, passes without any interruption  over the critical field value. (d), (e) Angular profiles of the magnetization for equally spaced values of the field, $h=0,\pm 0.1, \pm 0.2, \pm 03$ for both types of SkmLs. (f) Color plots of the magnetization at the verge of expansion at $h=\pm 0.3$ (upper and lower panels) as well as for $h=0$ (middle panel).
\label{fig:06}}
\end{figure*}

\section{Numerical solutions for isolated skyrmioniums}

In the present section, we validate the principles of skyrmionium internal stability drawn for the linear Ansatz. 
We obtain numerically rigorous solutions for isolated skyrmioniums by minimizing the functional (\ref{functional}) with non-dimensional units in MuMax3 \cite{mumax3}, i.e., the exchange and DMI constants are $A_{ex}=1, D_{DMI}=1$. The size of the numerical grid is $1024\times 1024$; the cell sizes are $0.1$ along all spatial directions. 
We construct the initial states for skyrmioniums by pinning the magnetization, $m_z=-1$, along the circle with the radius $R_{in}$ (we refer to this radius as $R_{in}^{pinned}$) and let MuMax3 relax the spin configuration.  Otherwise, the magnetization vectors point along $z$ axis, $m_z=1$. As it was argued in the previous section, no pinning is needed for the external radius $R_{ex}$. 

The energies of the internal skyrmion (the area within the circle with the radius $R_{in}$), external skyrmion (the area with $\rho>R_{in}$) and the skyrmionium are plotted in dependence on $R_{in}^{pinned}$ in Fig. \ref{fig:04} (a) with the same color coding as in Fig. \ref{fig:02} (a) and for the same control parameters, $h=0, k_u=0.7$.
The behavior is qualitatively the same as for the linear Ansatz, but with some nuances, which call for some additional explanation. 
First of all, we notice that for some critical radius $R_{in}^{pinned}$, a skyrmionium collapses into an ordinary skyrmion (the transition is highlighted by the dotted red line and the red arrow), i.e., the internal skyrmion becomes small within the given discretization scheme, and thus cell sizes must be reduced to address this type of skyrmionium solutions. The color plot for the $m_z$-component of the magnetization is shown in Fig. \ref{fig:05} (a). 
The problem of skyrmionium collapse into a skyrmion was recently addressed in Refs. \cite{Hagemeister,Jiang24} and thus will be omitted here.

For relatively large radii $R_{in}^{pinned}$, on the contrary, a skyrmionium transforms into a circular domain wall with comparable radii $R_{in}$ and $R_{ex}$, i.e., $\eta\rightarrow 1$ as was predicted by the linear Ansatz. The corresponding color plot is shown in Fig. \ref{fig:05} (b).

The energy of the external ring-shaped skyrmion (red curve in Fig. \ref{fig:04} (a)) gradually increases in dependence on $R_{in}^{pinned}$, and the total energy (green curve in Fig. \ref{fig:04} (a)) possesses a minimum for $R_{in}=5.4$. The internal skyrmion, however, exhibits an interesting evolution with the increasing $R_{in}^{pinned}$: once the energy of the internal skyrmion reaches the minimum, it's size stops growing, which is an artifact imposed by pinning.  To address this behavior, we introduce other characteristic radii $R_{in}^{Lilley}$ and $R_{ex}^{Lilley}$, which are defined according to the Lilley rule (Fig. \ref{fig:04} (b)):  we plot the second derivative $d^2m_z/d\rho^2$ (black curve), find its zero values and construct the tangent lines (green lines) to the profiles $m_z(\rho)$ (blue lines). Then, the intersection points with the magnetization levels $m_z=\pm 1$ constitute the required radii. 
All the scaffolding shown by the green lines in Fig. \ref{fig:04} (b) was done for the magnetization profile corresponding to the energy minimum. 
Dependences of both Lilley-radii are plotted in Fig. \ref{fig:04} (c). After the plateau with the constant $R_{in}^{Lilley}$, both radii continue to grow simultaneously, which is marked by discontinuities at all graphs. 
The color plots for the $m_z$-component of the magnetization, DMI and exchange energy densities as well as the total energy density are plotted in Fig. \ref{fig:05} (c) for the skyrmionium with the minimal total energy. All graphs exhibit characteristic target-like patterns and, e.g.,  allow to allocate three rings with the negative energy density and two rings with the positive one.

The ratio between two radii $\eta$ can be defined using both $R_{in,ex}^{Lilley}$ and $R_{in}^{pinned}$ (Fig. \ref{fig:04} (d)).  Although the ratio $R_{ex}^{Lilley}/R_{in}^{Lilley}$ (blue curve) exhibits discontinuity, both curve are consistent with the behavior of the linear Ansatz: for small radii $R_{in}$, the ratio $\eta$ increases; for large $R_{in}$, $\eta\rightarrow 1$.

\section{Skyrmionium lattices}

Skyrmionium "particles" may be also driven together to form skyrmionium lattices. The process of condensation is ruled by two competing mechanisms: low-energy skyrmionium cores trying to fill the whole space once they acquire the negative eigen-energy, and high-energy edge area responsible for the space frustration. 
On the contrary to SkLs, however, two varieties of SkmLs can be formed depending on the polarity of the central skyrmion. In the forthcoming simulations, we will consider just one type of SkmL, but will change the direction of the field. 

As foreseen by the linear Ansatz, both SkmLs represent metastable solutions, but with the metastability regions, which extend up to the critical line $e-c$ of the spiral state (red-shaded region in Fig. \ref{fig:06} (a)), i.e., SkmLs exist for the same control parameters as the spiral states, but bear higher energy.  

The lattice periods $p$ of both SkmLs (defined as the distances between the centers of two adjacent skyrmioniums) tend to infinity with approaching the critical line $e-c$. Fig. \ref{fig:06} (b) shows the period of the SkmL for the positive magnetic field and different values of the uniaxial anisotropy. The critical value of the field is essentially the same as for the spiral state. Indeed, such a SkmL transforms into a system of circular domain walls with the inflating radius (Fig. \ref{fig:06} (f), upper panel), i.e., on the contrary to the ordinary SkLs, the SkmL is not able to set free isolated skyrmioniums.  
For the negative magnetic field, the SkmL turns into a system of isolated skyrmions separated by the hexagonal web of narrow domain walls (Fig. \ref{fig:06} (f), bottom panel). The saturation field is slightly lower due to the small positive energy of the central skyrmions. Two-dimensional cross-cuts for both SkmLs are shown in Fig. \ref{fig:06} (d), (e) for different values of the applied magnetic field.

In addition, we plot the characteristic sizes  at some particular levels of the magnetization (Fig. \ref{fig:06} (c)): $\theta=3\pi/2$ (dark-blue curve), $\theta=\pi$ (green curve), and $\theta=\pi/2$ (light-blue curve), which also expand at the critical field values. The green curve signifies the width of the domain wall region, which is defined as $W=R_{\theta=3\pi/2}-R_{\theta=\pi/2}$. 

Interestingly, the characteristic sizes of isolated skyrmioniums also diverge while approaching the critical field from the side of higher fields (Fig. \ref{fig:06} (c)). The width $W$, however, preserves its continuity through the whole field range. The same behavior was observed for the ordinary SkL: also the period of the lattice diverges, the size of isolated skyrmions is finite for the critical field value. 
Within the used discretization scheme, isolated skyrmioniums were found to exist in the green-shaded region of the phase diagram (Fig. \ref{fig:06} (a)), and collapse above the line $k-l$. As was mentioned in Sect. \ref{section:Skmlinear}, smaller cell size of the numerical grids may lead to slightly higher values of the critical control parameters, which, however, are restricted by the physical arguments that the internal skyrmion cannot be infinitely small.

For small anisotropy values, some part of the area with isolated skyrmioniums is covered by the stable SkL \cite{Butenko}, which stabilizes below the line $m-e$ (dotted blue line in Fig. \ref{fig:06} (a)). For large anisotropy values, isolated skyrmioniums may exist alongside with the isolated skyrmions (to the right from the point $n$). To the right from the point $o$, the SkmL is energetically more favorable than SkL as was found for the spiral states.

\section{Conclusions}

In the present paper, we scrutinized the internal structure of skyrmioniums from the perspective of two communicating skyrmions (magnetic analogue of communicating water vessels). Within the central skyrmion, the magnetization performs its rotation according to the energy balance of the DMI and Zeeman/anisotropy contributions, which stipulates the energy minimum 
and resembles the behavior 
of an ordinary isolated skyrmion. The central skyrmion would dictate the same rotational algorithm within the surrounding ring-shaped skyrmion with the opposite polarity. This external skyrmion, however, does not exhibit any energy minimum, and would rather prefer to squeeze the central skyrmion to reduce its own positive energy. Thus, two constituent skyrmions within a skyrmionium communicate to establish an optimal ratio of their radii rather than each of the radii separately. 

For large values of the applied magnetic field and/or uniaxial anisotropy, skyrmioniums were shown to collapse. This process is accompanied by simultaneous shrinkage of both radii, but with the growth of their ratio, the process, which tries to prevent the skyrmionium destruction. Along another critical line at the phase diagram, 
both isolated skyrmioniums and skyrmionium lattices expand and transform into circular domain walls. Both skyrmionium realizations, however, exists as separate branches of solutions on the opposite sides from the critical line. Besides, skyrmionium lattices are overpowered by the 1D cycloids, which are energetically more preferable states and thus impede experimental realization of SkmLs in the whole  parameters region.

\section{ACKNOWLEDGMENTS}
The authors are grateful to Muneto Nitta and Yuki Amari for useful discussions.

\end{document}